\documentclass[
amsmath,
amssymb,
aps,
twocolumn,
]{revtex4-2}

\usepackage{graphicx}
\usepackage{dcolumn}
\usepackage{bm}
\usepackage{braket}
\usepackage{ulem}
\usepackage[version=3]{mhchem}
\usepackage{amsthm}
\usepackage{mathrsfs}
\usepackage{natbib}
\usepackage{url}
\usepackage[dvipdfmx,colorlinks=true,citecolor=blue,linkcolor=magenta]{hyperref}
\hypersetup{breaklinks=true}

\usepackage{color,ulem}

\newcommand{\dfigs}[1]{./#1}

\newcommand{\equref}[1]{(\ref{eq:#1})}
\newcommand{\figref}[1]{Fig.~\ref{fig:#1}}
\newcommand{\tabref}[1]{Tab.~\ref{tab:#1}}
\newcommand{\secref}[1]{Sec.~\ref{sec:#1}}

\newcommand{\Figref}[1]{Figure~\ref{fig:#1}}
\newcommand{\Tabref}[1]{Table~\ref{tab:#1}}
\newcommand{\Secref}[1]{Section~\ref{sec:#1}}

\newcommand{\mrm}{\mathrm}
\newcommand{\mbf}{\mathbf}
\newcommand{\mcl}{\mathcal}
\newcommand{\mbb}{\mathbb}

\newcommand{\br}[1]{\left( #1 \right)}
\newcommand{\Br}[1]{\left\{ #1 \right\}}
\newcommand{\BR}[1]{\left[ #1 \right]}

\newcommand{\av}[1]{\left| #1 \right|}

\newcommand{\hspc}{\hspace{1em}}

\newcommand{\uar}{\uparrow}
\newcommand{\dar}{\downarrow}

\newcommand{\dr}{\mrm{d}}
\newcommand{\pdr}{\mrm{\partial}}
\newcommand{\im}{\imath}
\newcommand{\ex}{e}
\newcommand{\alp}{\alpha}
\newcommand{\bet}{\beta}
\newcommand{\gam}{\gamma}
\newcommand{\del}{\delta}
\newcommand{\eps}{\epsilon}

\newcommand{\tht}{\theta}

\newcommand{\lam}{\lambda}
\newcommand{\sig}{\sigma}

\newcommand{\Del}{\Delta}

\newcommand{\Lam}{\Lambda}

\newcommand{\vphi}{\varphi}

\newcommand{\vtht}{\vartheta}
\newcommand{\ham}{\mcl{H}}
\newcommand{\pF}{\mcl{F}}
\newcommand{\pM}{\mcl{M}}
\newcommand{\pri}{\prime}

\newcommand{\rF}{\mrm{F}}
\newcommand{\rS}{\mrm{S}}
\newcommand{\rM}{\mrm{M}}
\newcommand{\Hc}{\mrm{H.c.}}
\newcommand{\gs}{\mrm{gs}}
\newcommand{\TB}{\mrm{TB}}

\begin{document}

\title{
Quantum-circuit algorithms for many-body topological invariant\\ and Majorana zero mode
}


\author{Takanori Sugimoto}
\email{sugimoto.takanori.qiqb@osaka-u.ac.jp}
\affiliation{Center for Quantum Information and Quantum Biology, Osaka University, Toyonaka, Osaka 560-8531, Japan} 
\affiliation{Advanced Science Research Center, Japan Atomic Energy Agency, Tokai, Ibaraki 319-1195, Japan}
\affiliation{Computational Materials Science Research Team, Riken Center for Computational Science (R-CCS), Kobe, Hyogo 650-0047, Japan}

\date{\today}

\begin{abstract}
  Topological states of matter are promising resources for composing fault-tolerant quantum computers, advancing beyond the limitations of current noisy intermediate-scale quantum devices. 
  To enable this progress, a deep understanding of topological phenomena within actual quantum computing platforms is essential.
  However, existing quantum-circuit algorithms to examine topological properties remain limited.
  Here we introduce three quantum-circuit algorithms designed to (i) determine the ground state within a specified parity subspace, (ii) identify the many-body topological invariant, and (iii) visualize zero-energy edge modes.
  We illustrate these algorithms with the interacting Kitaev chain, a typical model of one-dimensional topological superconductors.
  These approaches are versatile, extending beyond one-dimensional systems to various topological states, including those in higher dimensions.
\end{abstract}

\maketitle

\section{\label{sec:intro} Introduction}
Recent progress in quantum computing has raised expectations for achieving quantum supremacy, or at least, quantum advantage in the near future~\cite{Ladd2010,Xiang2013,Preskill2018,Harrow2017}.
Particularly, noisy intermediate-scale quantum (NISQ) devices based on the gate-type unitary operations are on the point of entering an unexplored region beyond the limit of numerical calculation that classical computers can approach in a feasible amount of time~\cite{Preskill2018,Boixo2018,Arute2019,Villalonga2020}.
Concurrently, numerous quantum-circuit (QC) algorithms, incorporating quantum gates and supplemented by auxiliary classical calculations, have been rapidly developed for various applications, such as the quantum approximate optimization algorithm~\cite{Farhi2014,Farhi2019,Hadfield2019,Zhou2020,Medvidovic2021,Wurtz2021,Wurtz2022,Yoshioka2023}, quantum Fourier transformation~\cite{Nielsen2000}, quantum singular-value decomposition~\cite{Rebentrost2018,Bravo-Prieto2020,Wang2021}, and quantum machine learning~\cite{Dunjko2016,Biamonte2017,Farhi2018,McClean2018,Mitarai2018,Perdomo-Ortiz2018,Cong2019,Havlicek2019,Huggins2019,Schuld2019}.

On the other hand, QC algorithms designed for condensed-matter physics are still limited.
Some fundamental algorithms have recently been proposed, such as the variational quantum eigensolver (VQE) for solving model Hamiltonians~\cite{Peruzzo2014,McClean2016,Kandala2017,Grimsley2019,Seki2020,Yordanov2020,Tang2021,Yordanov2021,Fujii2022,Mizuta2022,Anastasiou2022,Tilly2022} and algorithms for simulating dynamics and temperature dependence through real or imaginary time evolution~\cite{Wiebe2011,Smith2019,Barratt2021,Lin2021,Mizuta2023,Motta2020,Shirakawa2021,Miyakoshi2023}. 
However, algorithms specifically for analyzing topological properties are scarce~\cite{Cong2019,Smith2022,Xiao2022,Okada2022,Niedermeier2024}. 
To overcome coherence-time limitations and make strides toward fault-tolerant quantum computing (FTQC), it is crucial to obtain a deep understanding of topological states within quantum systems.
In this paper, we introduce a set of QC algorithms composed by three steps: (i) determining the ground state (GS) within a specified parity subspace, (ii) identifying the many-body topological invariant, and (iii) visualizing zero-energy edge modes.

For the demonstration of our algorithms, we utilize the interacting Kitaev chain, a model of a one-dimensional topological superconductor with many-body interactions~\cite{Kitaev2001,Fidkowski2010,Gangadharaiah2011,Sela2011,Hassler2012,Katsura2015,Miao2017}. 
The Kitaev chain is a well-known model within the BDI symmetry class, characterized by the conservation of time-reversal, particle-hole, and chiral (sublattice) symmetries~\cite{Schnyder2008,Fidkowski2011}.
Introducing many-body interactions in this model leads to a topological phase transition from a non-trivial to a trivial phase as defined by the topological invariant. 
Additionally, the topological superconducting phase of this model hosts zero-energy edge modes composed of Majorana fermions (MFs), known as Majorana zero modes (MZMs). 
Since MZMs are a promising resource for implementing braiding-based topological quantum computing, a key approach for future fault-tolerant quantum computing (FTQC)~\cite{Nayak2008,Sarma2015,Preskill2018}, visualizing their behavior is crucial from both foundational and engineering perspectives.

The rest of this paper is organized as follows.
In Sec.~II, we introduce the model Hamiltonian for the interacting Kitaev chain in the spinless-fermion representation, and outline its MF and spin counterparts.
In Sec.~III, as the first step of our QC algorithms, we apply the VQE technique to obtain the GS within the selected parity subspace and present the numerical results.
Sec.~IV provides an overview of the topological invariant in both the tight-binding (TB) and many-body (MB) models, representing systems without and with interaction, respectively. 
We then propose a QC algorithm to determine the MB topological invariant and show numerical results for various points in the model-parameter space, encompassing both topologically trivial and non-trivial states.
Additionally, in Sec.~V, we visualize the MZM by utilizing GSs from two distinct parity subspaces for the various points in the model-parameter space.
All numerical calculations were performed on the QC simulator, qulacs~\cite{Suzuki2021}, on a classical computer.
Finally, we conclude with a summary of our study, highlighting the advantages and limitations of our algorithms and addressing considerations for their application on real NISQ devices.

\section{\label{sec:model} Model}
In this section, we introduce the model Hamiltonian of one-dimensional topological superconductor, the so-called Kitaev chain with the attractive interaction on neighboring bonds~\cite{Kitaev2001,Sela2011,Hassler2012,Katsura2015,Miao2017}.
The model Hamiltonian for the $L$-site system is defined by,
\begin{align}
\label{eq:Kitaev}
&\ham_{\rF} =\notag\\
&-t\sum_{j=1}^{L-1} \br{c^\dag_j c_{j+1} + \Hc} -\Del\sum_{j=1}^{L-1} \br{c^\dag_j c_{j+1}^\dag + \Hc} \notag \\
&- V \sum_{j=1}^{L-1} \br{n_j-\frac{1}{2}} \br{n_{j+1}-\frac{1}{2}} -\mu\sum_{j=1}^L  \br{n_{j}-\frac{1}{2}} \notag\\
&- B \br{t\,c^\dag_L c_{1} + \Del\, c^\dag_L c_{1}^\dag + \Hc} - B^\pri V \br{n_L-\frac{1}{2}} \br{n_{1}-\frac{1}{2}} 
\end{align}
where $c_j$, $c_j^\dag$, and $n_j$ denote annihilation, creation, and number operators of spinless fermion at $j$th site, respectively.
In addition, the hopping integral, the superconducting pairing potential, and the Coulomb potential between neighboring sites are given by $t$, $\Del$, and $V$, respectively, with the chemical potential $\mu$.
The boundary conditions (BCs) are introduced by $B$ and $B^\pri$: $B=B^\pri=0$ for the open boundary condition (OBC), $B=B^\pri=1$ for the periodic boundary condition (PBC), and $-B=B^\pri=1$ for the antiperiodic boundary condition (ABC).
In this paper, we focus on the attractive region for the Coulomb potential $V>0$, to avoid the trivial phase of the repulsive region, where the translational symmetry is spontaneously broken~\cite{Hassler2012,Katsura2015,Miao2017}.
Without the interaction, we can easily understand the topological invariant and the MZM, based on the one-particle picture (see \Secref{top} for the topological invariant and \Secref{MZM} for the MZM).
Since the Kitaev chain has pair creation and annihilation terms [$\Del$ terms in \equref{Kitaev}], the total number of fermions $N=\sum_jn_j$ is not conserved.
Instead, the fermion parity $\pF=\exp[\im\pi\sum_jn_j]=\pm 1$, i.e., whether the number of fermions is even or odd, is a good quantum number.

\begin{figure}[htb]
  \includegraphics[width=0.85\columnwidth]{\dfigs{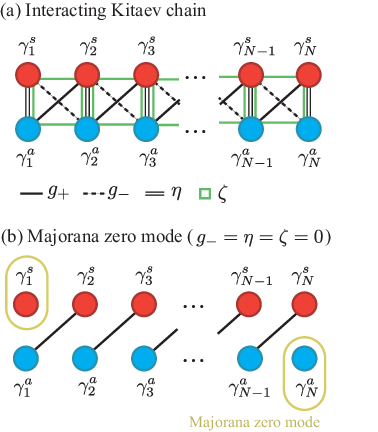}}
  \caption{(a) Schematic of the interacting Kitaev chain with OBC ($B=B^\pri=0$) in the MF representation \equref{MFH}. Four coupling terms, $g_+$, $g_-$, $\eta$, and $\zeta$ are denoted by solid lines, dashed lines, and double solid lines, and green squares, respectively. Red (blue) balls are the symmetric (anti-symmetric) modes of the MFs. (b) Majorana zero mode (MZM) for $g_-=\eta=\zeta=0$. Two MFs $\gam_1^s$ and $\gam_L^a$ at the edges are decoupled from the system, corresponding to the MZM.}
  \label{fig:model}
\end{figure}

\begin{table}
  \caption{Correspondence of the BCs for two subspaces of the fermion parity $\pF=\pm1$ between the Kitaev chain and the spin chain.}
  \begin{tabular}{c|rr}
    BCs  & Kitaev chain ($\ham_\rF$) & spin chain ($\ham_\rS$) \\ \hline\hline
    $B=B^\pri=0$ & OBC ($\pF=\pm1$) & OBC ($\pF=\pm1$) \\
    $B=B^\pri=1$ & PBC ($\pF=\pm1$) & $ \begin{cases} \mrm{ABC} & (\pF=+1) \\ \mrm{PBC} & (\pF=-1) \end{cases}$\\
    $-B=B^\pri=1$ & ABC ($\pF=\pm1$) & $ \begin{cases} \mrm{PBC} & (\pF=+1) \\ \mrm{ABC} & (\pF=-1) \end{cases}$
  \end{tabular}
  \label{tab:BC}
\end{table}

The Kitaev chain mathematically corresponds to the $S=1/2$ XYZ spin chain,
\begin{align}
\label{eq:XYZ} 
\ham_{\rS} =& -\sum_{\alp=x,y,z} J_\alp \sum_{j=1}^{L-1} S_j^\alp S_{j+1}^\alp - h_z\sum_{j=1}^L S_j^z \notag\\
&+ B \sum_{\alp=x,y} J_\alp \pF S_L^\alp S_{1}^\alp - B^\pri J_z S_L^z S_{1}^z,
\end{align}
via the Jordan--Wigner (JW) transformation,
\begin{equation}
c_j = S_j^- \ex^{\im \vphi_j}, \ c_j^\dag = S_j^+ \ex^{\im \vphi_j}, \ n_j = S_j^z +\frac{1}{2}.
\end{equation}
The JW phase is defined by $\vphi_j=\pi\sum_{i=1}^{j-1}(S_i^z +\frac{1}{2})$ with the imaginary unit $\im=\sqrt{-1}$, and $S_j^\alp$ ($S_j^\pm$) represents the $\alp=x,y,z$ component of $S=1/2$ spin operator (the ladder operator of spin) at $j$th site with the natural unit $\hbar=1$.
The anisotropic exchange interaction is denoted by $J_\alp$, and $h_z$ represents the magnetic field along $z$ axis.
The coupling terms in the Kitaev chain and the XYZ spin chain have the following relations:
\begin{equation}
t=\frac{J_x+J_y}{4},\ \Del=\frac{J_x-J_y}{4},\ V=J_z, \ \mu=h_z.
\end{equation}
Since the QC is compatible to the spin representation, we mainly use the spin representation of the Hamiltonian in this paper.
However, it is noteworthy that careful consideration is required to the fermion parity of the subspace on which the spin Hamiltonian acts, because the boundary terms of the spin Hamiltonian include the fermion parity $\pF$.
\Tabref{BC} shows the relationship of the three BCs for two subspaces of the fermion parity $\pF$ between the Kitaev chain and the spin chain.
The PBC in the Kitaev chain corresponds to the ABC for $\pF=+1$ and the PBC for $\pF=+1$ in the spin chain.

To understand topological properties in the Kitaev chain, the MF representation is also important.
The MF representation of the Kitaev chain \equref{Kitaev} is given by,
\begin{align}
\label{eq:MFH}
\ham_{\rM} =\, &-\im g_- \sum_{j=1}^{L-1} \gam_j^s \gam_{j+1}^a + \im g_+ \sum_{j=1}
^{L-1} \gam_j^a \gam_{j+1}^s \notag\\
&+ \zeta \sum_{j=1}^{L-1} \gam_j^s \gam_j^a \gam_{j+1}^s\gam_{j+1}^a -\im\eta \sum_{j=1}^L \gam_j^s \gam_j^a\notag\\
&-\im B \br{g_- \gam_L^s \gam_{1}^a - g_+ \gam_L^a \gam_{1}^s }+ B^\pri \zeta \gam_L^s \gam_L^a \gam_{1}^s\gam_{1}^a
\end{align}
with the coupling constants $g_\pm=(t\pm\Del)/2$, $\zeta=V/4$, and $\eta=\mu/2$.
The symmetric ($\gam_j^s$) and antisymmetric ($\gam_j^a$) modes of MF at $j$th site are defined by,
\begin{equation}
\label{eq:Majorana-ops}
\gamma_j^s = c_j^\dag + c_j, \quad \gamma_j^a =  \im\br{c_j^\dag - c_j}.
\end{equation}
Note that the MF operators obey the fermionic anti-commutation relation $\{ \gam_j^\tau , \gam_{j^\pri}^{\tau^\pri} \} = 2\del_{j,j^\pri}\del_{\tau,\tau^\pri}$ with the Hermiticity $(\gam_j^\tau)^\dag=\gam_j^\tau$ for $\tau (\tau^\pri)=s,a$.

\Figref{model}(a) shows the schematic of the MF Hamiltonian~\equref{MFH} with OBC ($B=B^\pri=0$).
A pair of symmetric (red ball) and antisymmetric (blue ball) MFs corresponds to a spinless fermion.
When $g_-=\eta=\zeta=0$, the $g_+$ diagonal coupling only remains, so that two MFs $\gam_1^s$ and $\gam_L^a$ at the edges are decoupled from the system.
The pair of the MFs are regarded as a spinless fermion with zero energy (the MZMs), inducing two-fold degeneracy at every energy level.
The existence of the MZMs is also an evidence of topological superconductor.

\section{\label{sec:GS} Parity-selected GS}
Next, we explain how to obtain the GSs utilizing a QC algorithm with conserving the fermion parity.
The fermion parity is related to the magnetization parity in the XYZ spin chain, $\pF=\im^L\pM_z$ with
\begin{equation}
  \pM_\alp=\exp\BR{\im\pi\sum_j S_j^\alp}=\im^L \prod_j (2S_j^\alp),
\end{equation}
for $\alp=x,y,z$.
Hence, the $\pM_z$ is also a good quantum number, corresponding to the fermion parity when $L=0$ (mod. $4$).
It is worth noting that the other components of magnetization parity $\pM_x$ and $\pM_y$ are conserved, while the three magnetization parities are not independent due to the relation $\pM_x\pM_y\pM_z=\im^{3L} \prod_j (8S_j^x S_j^y S_j^z)=1$~\cite{Wada2021}.
In this paper, to avoid misunderstanding due to the system-size dependence of the fermion parity, we consider only the system sizes satisfying $L=0$ (mod. $4$).

\subsection{GS energies with several BCs}

\begin{figure}[tb]
  \includegraphics[width=0.93\columnwidth]{\dfigs{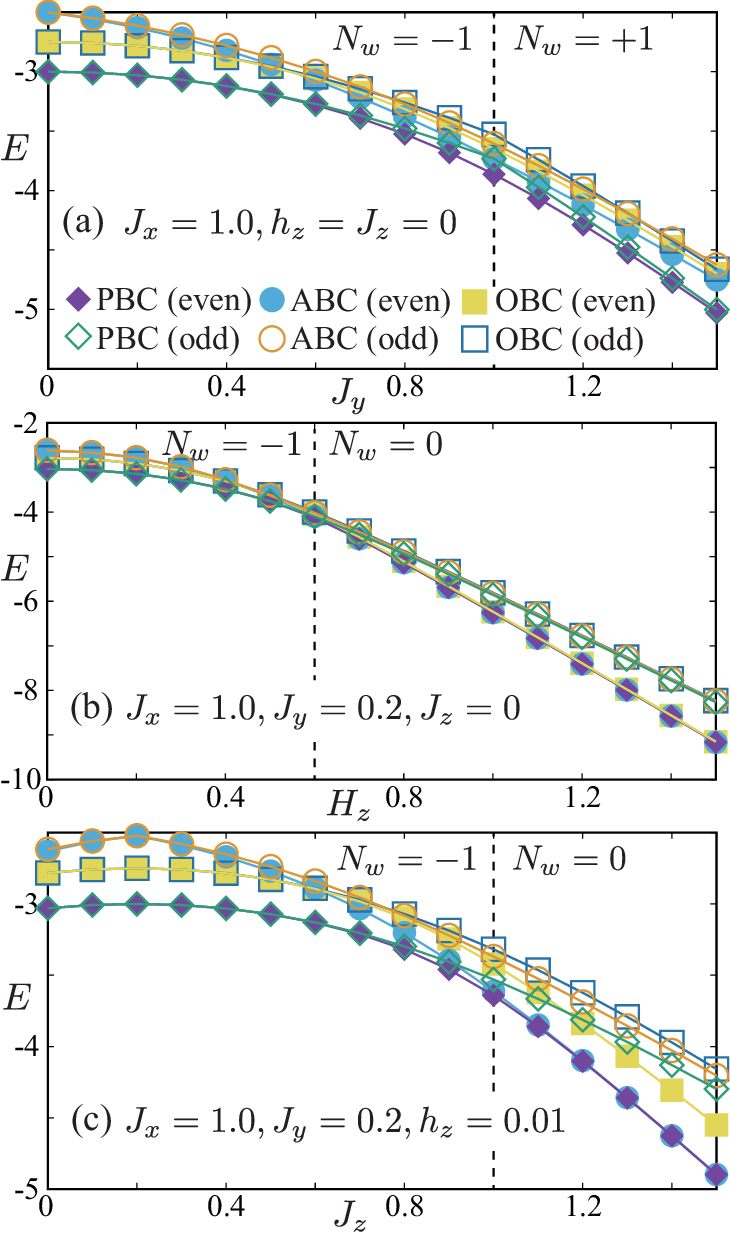}}
  \caption{GS energies of the 12-site XYZ spin chain with PBC (diamond), ABC (circle), and OBC (square), obtained by the ED method. The even (odd) parity of $\pF$, i.e., $\pF=+1$ ($\pF=-1$), is denoted by close (open) symbol. See \tabref{BC} for the correspondence of the BCs and the parity between the XYZ spin chain and the Kitaev chain. (a) $J_y$ dependence with fixed $h_z=J_z=0$. (b) $h_z$ dependence with fixed $J_y=0.2$ and $h_z=0$. (c) $J_z$ dependence with fixed $J_y=0.2$ and $h_z=0.01$. $J_x$ is set to the energy unit. To avoid degeneracy with large $J_z$ in topologically-trivial state, we add a small magnetic field $h_z=0.01$. $N_w$ represents the winding number as the topological invariant of the Kitaev chain (explained in \secref{top}). The vertical dashed lines denote the topological phase transition points in the thermodynamic limit~\cite{Hassler2012}.}  
  \label{fig:ed}
\end{figure}

Before the QC algorithm for topological features of the XYZ spin chain, we briefly explain level-crossing behaviors and finite-size effects in GS energies with several BCs in two parity subspaces.
\Figref{ed} shows the GS energies of the 12-site XYZ spin chain obtained by exact diagonalization (ED) method.  
Three panels of \figref{ed} represent $J_y$, $h_z$, and $J_z$ dependence, respectively, with fixed other parameters.
Based on \tabref{BC}, the Kitaev chain with PBC for $\pF=\pm1$ (ABC for $\pF=\pm1$) corresponds to the spin chain with ABC for $\pF=+1$ and PBC for $\pF=-1$ (PBC for $\pF=+1$ and ABC for $\pF=-1$), respectively.
In \figref{ed}(a), the energy of PBC with odd parity is lower than that of ABC with even parity, while in \figref{ed}(b) and (c), we can see the level-crossing between them near the dashed lines.
On the other hand, the energy of PBC with even parity is always lower than that of ABC with odd parity. 
The GS is topological when the product of GS parities with PBC and ABC in the Kitaev chain is odd~\cite{Kawabata2017}.
Namely, the level-crossing between the GS energies in  \figref{ed}(b,c) indicates the topological-to-trivial phase transition.
Hence, we can clarify whether the topological state emerges or not, by examining GS parities with PBC and ABC utilizing a parity-selected VQE method which can determine the GS energy within a specified parity subspace (explained below). 
However, topological transitions between different topological invariants in \figref{ed}(a) is not detected by examining GS parities.
Moreover, it is noteworthy that even in the topological phase, the degeneracy of the GS energies with OBC caused by the topological edge state [see \figref{model}(b)] is not exactly obtained because of the finite-size effects.
Consequently, in addition to the parity-selected VQE method, a quantum algorithm directly determining the topological invariant is also necessary.

\subsection{Parity-selected VQE method}

\begin{figure*}[tb]
  \includegraphics[width=0.9\textwidth]{\dfigs{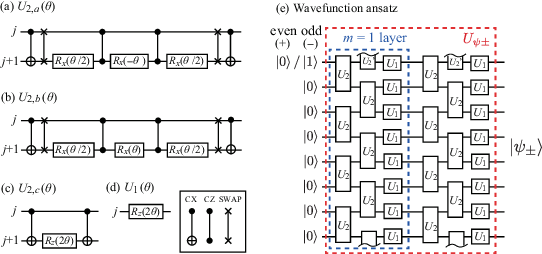}}
  \caption{(a-d) QC representation of three unitary operators $U_{2,p}(\tht)$ for $p=a,b,c$ and $U_1(\tht)$, where one-qubit rotation around $\alp=x,y,z$ axis is denoted by $R_\alp(\tht)=\exp[\im (\tht/2) \sig^\alp]$ with the Pauli matrix $\sig^\alp$. (e) Wavefunction ansatz for the parity-selected VQE method ($L=8$, $M=2$). We adopt a brick wall type of QC constructed by the $U_{2}(\bm{\tht}) = U_{2,a}(\tht_{a})U_{2,b}(\tht_{b})U_{2,c}(\tht_{c})$ and $U_1(\vtht)$ gates with $\bm{\tht}=(\tht_a,\tht_b,\tht_c)$. The angles $\bm{\tht}$ and $\vtht$ are assigned to variational parameters for the VQE optimization. The 2-qubit unitary operators bridging on the edges are set to the identity operator for OBC [$U_{2}^\pri(\bm{\tht})=1$] and the rotation operator for PBC and ABC [$U_{2}^\pri(\bm{\tht})=U_{2}(\bm{\tht})$].  Thus, the total number of variational parameters is $N_\tht=(4L-3)M$ for OBC and $N_\tht=4LM$ for PBC and ABC.}
  \label{fig:ugate}
\end{figure*}

To implement the parity-selected GSs into QCs, we extend the VQE method to restricted Hilbert subspaces, which we call parity-selected VQE method.
The parity-selected VQE method is based on wavefunction ansatz consisting of parity-conserved unitary gates and initial states as eigenstates of the parity operator. 
For the XYZ spin chain with the magnetization parity $\pM_z$, we use the parity-conserved 2-qubit unitary operator on $j$th bond defined by
\begin{equation}
\label{eq:u2}
U_{2}(\bm{\tht}) = U_{2,a}(\tht_{a})U_{2,b}(\tht_{b})U_{2,c}(\tht_{c})
\end{equation}
with
\begin{align}
U_{2,a}(\tht) &= \exp\BR{2\im\tht \br{S_j^xS_{j+1}^x+S_j^yS_{j+1}^y}}\notag\\
&=\exp\BR{\im\tht\br{S_j^+ S_{j+1}^- +\Hc }},\\
U_{2,b}(\tht) &= \exp\BR{2\im\tht \br{S_j^xS_{j+1}^x-S_j^yS_{j+1}^y} }\notag\\
&=\exp\BR{\im\tht\br{S_j^+ S_{j+1}^+ +\Hc }},\\
U_{2,c}(\tht) &= \exp\BR{4\im\tht S_j^zS_{j+1}^z},
\end{align}
where the vector of angles includes three angles, $\bm{\tht}=(\tht_a,\tht_b,\tht_c)$.
Note that these operators $U_{2,p}(\tht)$ ($p=a,b,c$) on the $j$th bond, commute with each other, $[U_{2,p}(\tht),U_{2,p^\pri}(\tht^\pri)]=0$, while the unitary operators do not always commute between the neighboring bonds.
In addition, to take into account the effect of magnetic field, we introduce the 1-site unitary operator at $j$th site given by
\begin{equation}
\label{eq:u1}
U_{1}(\tht) = R_z(2\tht)= \exp\BR{2\im\tht S_j^z}.
\end{equation}
Although the unitary operators $U_{2}$ and $U_{1}$ have the site dependence, we omit the site index $j$ in the notation of unitary operators \equref{u2} and \equref{u1} for simplicity.

\begin{figure*}[tb]
  \includegraphics[width=1.0\textwidth]{\dfigs{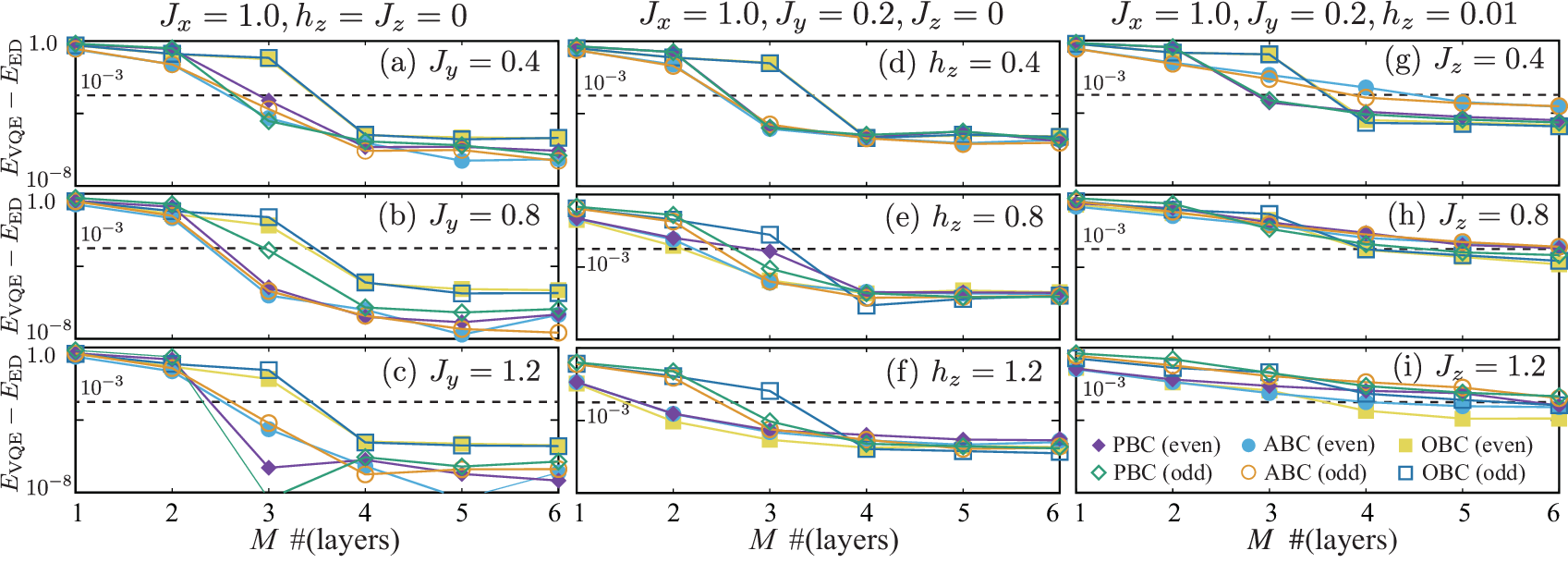}}
  \caption{$M$ dependence of GS energy differences between the parity-selected VQE and ED methods in the 12-site XYZ spin chain for various model parameters: (a-c) $J_y$ dependence with fixed $h_z=J_z=0$, (d-f) $h_z$ dependence with fixed $J_y=0.2$ and $J_z=0$, and (g-i) $J_z$ dependence with fixed $J_y=0.2$ and $h_z=0.01$. The GS energy of the parity-selected VQE method is the minimal energy in 10 times trials of optimization from random initial angles. The GS energies obtained by the ED method and the topological phase boundaries are shown in \figref{ed}. The symbols denote the BCs and the parities, corresponding to the symbols in \figref{ed}. The horizontal dashed line corresponds to $10^{-3}$.}  
  \label{fig:den}
\end{figure*}

\Figref{ugate} shows the QC representation of these unitary operators.
Since these operators preserve the fermion parity $[U_{2,p}(\tht),\pF]=[U_{1}(\vtht),\pF]=0$, the fermion parity after the unitary operations equals the parity of the initial state.
By using the parity-conserved 2-qubit and 1-qubit unitary operators, we adopt the following wavefunction ansatz for the parity-selected VQE method:
\begin{align}
&\ket{\psi_\pm(\{\bm{\tht}_{j,m}, \vtht_{j,m}\})} =U_{\psi}^\pm\ket{\mrm{i}_\pm},
\end{align}
with
\begin{equation}
U_{\psi}^\pm=\prod_{m=1}^{M}\BR{\prod_{j} U_{1}(\vtht_{j,m})\prod_{\mrm{even}\,j} U_{2}(\bm{\tht}_{j,m}) \prod_{\mrm{odd}\,j} U_{2}(\bm{\tht}_{j,m})}
\end{equation}
and the initial state for even parity $\ket{\mrm{i}_+}=\ket{0}^{\otimes L}$ or odd parity $\ket{\mrm{i}_-}=\sig_1^x\ket{\mrm{i}_+}=\ket{1}\otimes\ket{0}^{\otimes L-1}$.
$M$ is the number of layers.
The total number of variational parameters (namely, the number of angles in unitary operators $\{\bm{\tht}_{j,m}\}$ and $\{\vtht_{j,m}\}$) corresponds to $N_\tht=(4L-3)M$ for OBC and $N_\tht=4LM$ for PBC and ABC.
With these angles, we perform the parity-selected VQE calculation, i.e., optimization of the angles to minimize the expectation value of energy for the Hamiltonian~\equref{XYZ},
\begin{equation}
E_\pm(\Br{\bm{\tht}_{j,m},\vtht_{j,m}})=\bra{\psi_\pm} \ham_\rS \ket{\psi_\pm}
\end{equation}
with the quantum simulator, qulacs~\cite{Suzuki2021}, in the classical computer.

As the optimization method, we use the (dual) simulated annealing (SA) and Broyden--Fletcher--Goldfarb--Shanno (BFGS) algorithms served by the python library SciPy~\cite{Virtanen2020}.
The SA calculation is used to prepare appropriate initial angles $\{\bm{\tht}_{j,m},\vtht_{j,m}\}$ for the BFGS calculation, avoiding the local minima.
Nevertheless, we sometimes failed to obtain the global energy minimum, so that we regard the minimal-energy state in 10 times trials including both the SA and BFGS optimizations starting with random initial angles, as the GS.
For optimization of the BFGS method, we set the acceptable error $\Del E<10^{-8}$, where $\Del E$ is the energy difference between current and previous iterations in the main loop of the BFGS method.

\Figref{den} shows the number of layers ($M$) dependence of GS energy differences between the parity-selected VQE and ED methods for various points in the model-parameter space of the 12-site XYZ spin chain.
The left, center, and right panels in \figref{den} show the XY anisotropy ($J_y$), the magnetic field ($h_z$), and the Ising term ($J_z$) dependencies, respectively, with fixed other parameters, corresponding to those in \figref{ed}.
Firstly, we can see the good convergence of the GS energies obtained by the parity-selected VQE method; the energy differences from the ED method are almost smaller than $10^{-3}$.
However, the energy difference with odd parity (open symbol) is basically larger than even parity (close symbol) for each BC.
The reason might be that the initial state before unitary operations in the QC (i.e., $\ket{\mrm{i}_\pm}$) is not uniform with odd parity but with even parity.
Additionally, the energy convergence with large $J_z$ [\figref{den}(g-i)] looks worse, as compared with other parameters.  
The GS with large $J_z$ is the so-called Schr\"odinger's cat state~\cite{Miao2017}, which is the superposition of macroscopic classical state like $(\ket{\uar\uar\cdots\uar}\pm\ket{\dar\dar\cdots\dar})/\sqrt{2}$.
Since the Schr\"odinger's cat state may require many swap operations, the energy difference is worse with large $J_z$ in the shallow QCs.
However, the verification is basically out of scope in this paper, because the large $J_z$ region is the topologically-trivial phase, so that we leave it to future research.

\section{\label{sec:top} Topological invariant}
In this section, we introduce the MB topological invariant after a brief explanation of the TB topological invariant.
The MB topological invariant is an extension of topological invariant determined by one-particle picture, that is, the TB model.
Hence, we start with non-interacting Kitaev chain ($V=0$) with PBC as the TB model.

\subsection{Tight-binding (TB) model}
The TB model of the Kitaev chain with PBC is defined by,
\begin{equation}
\ham_\rF^{(\TB)} = \ham_\rF|_{V=0,B=B^\pri=1} .
\end{equation}
The Fourier transform $c_k=L^{-1/2}\sum_k c_j \ex^{\im jk}$ to the wavenumber $k=2\pi l/L$ ($l=-L/2,-L/2+1,\cdots,L/2-1$ for even $L$) gives the momentum-space Hamiltonian,
\begin{align}
\ham_\rF^{(\TB)} &= \sum_k \BR{2\eps_k c_k^\dag c_k + (-\im\Del_k c_k^\dag c_{-k}^\dag +\Hc)} \notag\\
&=\sum_k \bm{c}_k^\dag \mbf{H}_k \bm{c}_k -\mu L/2,
\end{align}
with $\eps_k=-t\cos k-\mu/2$ and $\Del_k=-\Del \sin k$.
The matrix form is the Nambu representation of the Hamiltonian given by,
\begin{equation}
\mbf{H}_k=\begin{pmatrix}
\eps_k & -\im \Del_k\\
\im \Del_k & -\eps_k
\end{pmatrix},\ \bm{c}_k=\begin{pmatrix}
c_k \\ c_{-k}^\dag
\end{pmatrix}.
\end{equation}

Since the coupling matrix is rewritten by $\mbf{H}_k=\eps_k\sig^z+\Del_k\sig^y=\bm{v}_\mrm{A}\cdot \bm{\sig}$ with the so-called Anderson pseudo vector $\bm{v}_\mrm{A}=(0,\Del_k,\eps_k)$ and the Pauli matrices $\sig^\alp$ ($\alp=x,y,z$), the rotation around $x$ axis $R_x(\phi)$ can diagonalize the coupling matrix.
By setting the rotation angle to $\phi_k=\tan^{-1}(\Del_k/\eps_k)$, we obtain the diagonalized Hamiltonian,
\begin{equation}
\label{eq:Kitaev-tb}
\ham_\rF^{(\TB)}= 2\sum_k \xi_k \bet_k^\dag \bet_k +E_\gs,
\end{equation}
with the dispersion relation $\xi_k=\sqrt{\eps_k^2+\Del_k^2}$, where the GS energy
\begin{equation}
\label{eq:egs-tb}
 E_\gs^{(\TB)}= -\sum_k \br{\xi_k+\mu/2}
\end{equation}
and the bogolon operator
\begin{equation}
\bet_k=\cos(\phi_k/2) c_k-\im \sin(\phi_k/2) c_{-k}^\dag.
\end{equation}
Since the Hamiltonian is diagonalized by the bogolon operator, the GS is the vacuum of the bogolon,
\begin{equation}
\label{eq:gs-tb}
\ket{\gs}_{\TB}=\prod_{0<k<\pi} \BR{\cos\br{\frac{\phi_k}{2}}+\im\sin\br{\frac{\phi_k}{2}}c_k^\dag c_{-k}^\dag }c_0^\dag\ket{0}.
\end{equation}

The topological invariant of the TB Kitaev chain \equref{Kitaev-tb}, the so-called winding number, is defined by
\begin{equation}
\label{eq:wn-tb}
N_w^{(\TB)}=\frac{1}{2\pi}\int_{k=-\pi}^\pi \dr \phi_k.
\end{equation}
The topological invariant represents the number of times that the Anderson pseudo vector $\bm{v}_\mrm{A}$ circulates counterclockwise around $x$ axis.
Thus, the topological phase appears when $|\mu|<2t$ ($t>0$) with finite pairing potential $\Del\neq 0$.
The sign of the pairing potential affects the sign of the winding number; $N_w=-1$ ($N_w=1$) for $\Del>0$ ($\Del<0$) if $t>0$ and $|\mu|<2t$.

\subsection{Many-body (MB) model}
In the TB model, since the electron with wavenumber $k$ interacts only with the electron of wavenumber $-k$, we can clearly write down the GS in the momentum space, and calculate the winding number, as a topological invariant defined by the coefficients of momentum-space GS.
However, if the interaction $V$ is introduced, the one-particle representation of the GS is difficult to obtain analytically in general~\footnote{Note that the exact solution is shown in the special case~\cite{Katsura2015,Miao2017}.}, because the interaction hybridizes all electrons with any momentum,
\begin{equation}
\label{eq:int-pbc}
V \br{\sum_{j=1}^{L-1} n_j n_{j+1} + n_L n_1} = \frac{V}{L}\sum_{q,k,k^\pri}c_k^\dag c_{k+q} c_{k^\pri}^\dag c_{k^\pri-q}.
\end{equation}
Hence, the GS is not the direct-product form of the bogolon wavefunctions.
Instead of the TB winding number \equref{wn-tb}, we adopt the MB winding number~\cite{Gurarie2011,Manmana2012,Li2018} given by,
\begin{equation}
\label{eq:wn}
N_w=\frac{1}{4\pi\im}\int_{-\pi}^\pi \dr k\, \mrm{Tr}\BR{\sig^x \mbf{G}_k^{-1} \pdr_k \mbf{G}_k}
\end{equation}
where $\mbf{G}_k$ represents the $2\times 2$ matrix of the Green functions of zero frequency,
\begin{equation}
\mbf{G}_k=\begin{pmatrix}
g_{c_k^\dag,c_k} & g_{c_{-k},c_k} \\
g_{c_k^\dag,c_{-k}^\dag} & g_{c_{-k},c_{-k}^\dag}
\end{pmatrix}
\end{equation}
with
\begin{equation}
g_{A,B}=\bra{\gs}A\frac{1}{\ham_\rF-E_{\gs}}B\ket{\gs} - \bra{\gs}B\frac{1}{\ham_\rF-E_{\gs}}A\ket{\gs}.
\end{equation}

For instance, by using \equref{Kitaev-tb},\equref{egs-tb} and \equref{gs-tb}, we can obtain the matrix $\mbf{G}_k$ in  the TB model as
\begin{equation}
\label{eq:gmat-tb}
\mbf{G}_k^{(\TB)}=\frac{1}{\sqrt{\eps_k^2+\Del_k^2}}\begin{pmatrix}
-\cos\phi_k & \im\sin\phi_k \\
-\im\sin\phi_k & \cos\phi_k
\end{pmatrix}
=\frac{-\mbf{H}_k}{\eps_k^2+\Del_k^2}.
\end{equation}
Therefore, we can confirm that the MB winding number \equref{wn} for the TB model corresponds to the TB winding number \equref{wn-tb},
\begin{align}
N_w&=\frac{1}{4\pi\im}\int_{-\pi}^\pi \dr k\, \mrm{Tr}\BR{\sig^x \Br{\mbf{G}_k^{(\TB)}}^{-1} \pdr_k \mbf{G}_k^{(\TB)}}\notag\\
&=\frac{1}{4\pi\im}\int_{-\pi}^\pi \dr k\, (2\im \pdr_k \phi_k)=N_w^{(\TB)}.
\end{align}
In the Green-function matrix for the TB model \equref{gmat-tb}, we can see the relations between the matrix elements: $g_{c_k^\dag,c_k}=-g_{c_{-k},c_{-k}^\dag}\in \mbb{R}$ and $g_{c_{-k},c_k}=-g_{c_{k}^\dag,c_{-k}^\dag}\in \im\mbb{R}$.
These relations are reserved even with the MB interaction \equref{int-pbc}, because the time-reversal symmetry ($\mcl{T}:c_k\to c_{-k},\ \im\to -\im$) protecting the relations, is kept.
Based on the relations, we can simplify the MB winding number as follows,
\begin{equation}
N_w=\frac{1}{2\pi\im}\int_{k=-\pi}^\pi \dr \log Z_k
\label{eq:MBW}
\end{equation}
with the MB counterpart of the Anderson pseudo vector in the complex plane,
\begin{equation}
\label{eq:gf-mf}
Z_k=-\frac{\im}{2}g_{c_k^\dag+c_{-k},c_k-c_{-k}^\dag}=\frac{1}{2}g_{\gam_k^s,\gam_{-k}^a},
\end{equation}
where the Fourier transform of MFs is defined by $\gam_k^\tau=L^{-1/2}\sum_k \gam_j^\tau\ex^{-\im jk}$ ($\tau=s,a$).

\begin{figure}[tb]
\includegraphics[width=0.9\columnwidth]{\dfigs{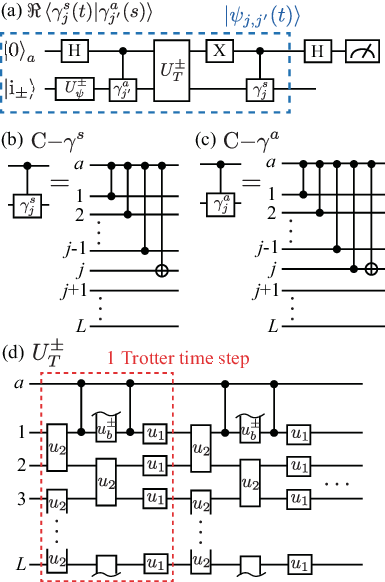}}
\caption{(a) QC algorithm of $\Re \braket{\gam_j^s(t)|\gam_{j^\pri}^a(t)}$ based on the Hadamard test. The real GS is chosen between the GSs with different parities, $\ket{\mrm{gs}_\pm}=U_\psi^\pm\ket{\mrm{i}_\pm}$, where $U_\psi^\pm$ represents the unitary operator obtained by the parity-selected VQE optimization. (b,c) Control-$\gam$ gates used in (a), consisting of CZ and CX gates. (d) Unitary gate of time evolution $U_T^\pm$, consisting of the Trotter decomposition of $\ex^{-\im \ham_S t}$ into the product of the infinitesimal time evolutions with local unitary operators, $u_2(\Del t)$, $u_1(\Del t)$, and $u_b^\pm(\Del t)$ (see the main text for the definition).}
\label{fig:top}
\end{figure}

\subsection{QC algorithm}
In the finite-size systems, the MB winding number \equref{MBW} are discretized as follows,
\begin{equation}
   N_w=\frac{1}{2\pi}\sum_{k} \Im\log \BR{Z_{k+\Del k} Z_k^\ast},
\end{equation}
with $\Del k=2\pi/L$.
Additionally, to determine the MB Anderson pseudo vector in the QC, we need to calculate the Green functions of MFs \equref{gf-mf} in the real space,
\begin{equation}
Z_k=\frac{1}{2}g_{\gam_k^s,\gam_{-k}^a}=\frac{1}{2L}\sum_{j,j^\pri}\ex^{-\im (j-j^\pri) k} g_{\gam_j^s,\gam_{j^\pri}^a}.
\end{equation}
To obtain the real-space Green function of MFs, we rewrite it by using the time-evolution form:
\begin{align}
  \label{eq:TEGF}
g_{\gam_j^s,\gam_{j^\pri}^a}&=2\Im \bra{\gs}\gam_j^s\frac{1}{\ham_\rF-E_{\gs}}\gam_{j^\pri}^a\ket{\gs} \notag\\
&=-\lim_{\del\to+0}\lim_{T\to\infty}2\int_0^T\dr t\, \ex^{-\del t}  \Re \braket{\gam_j^s(t)|\gam_{j^\pri}^a(t)},
\end{align}
with two time-evolved MF-excited states,
\begin{equation}
  \label{eq:TEMFES}
\ket{\gam_j^s(t)}=\gam_j^s\ex^{-\im \ham_\rF t}\ket{\gs},\ \ket{\gam_{j}^a(t)}=\ex^{-\im \ham_\rF t}\gam_{j}^a\ket{\gs}.
\end{equation}
Here, although we introduce the infinitesimal damping factor $\del\to+0$ and the infinite cutoff time $T\to\infty$, these are set to be finite values in the numerical calculation.
We should set the cutoff time $T$, satisfying $T\del\gg 1$, with the small enough damping factor $\del\ll 1$.
The effects of finite values are important to obtain the winding number by using the Green-function matrix, whereas the effects are not clarified so far.
Thus, the damping-factor dependence of $Z_k$ in the numerical calculations is discussed below.
In addition, we have to take care of the BCs in the XYZ spin chain, corresponding to the Kitaev chain $\ham_\rF$ with PBC.

\begin{figure*}[tb]
\includegraphics[width=0.95\textwidth]{\dfigs{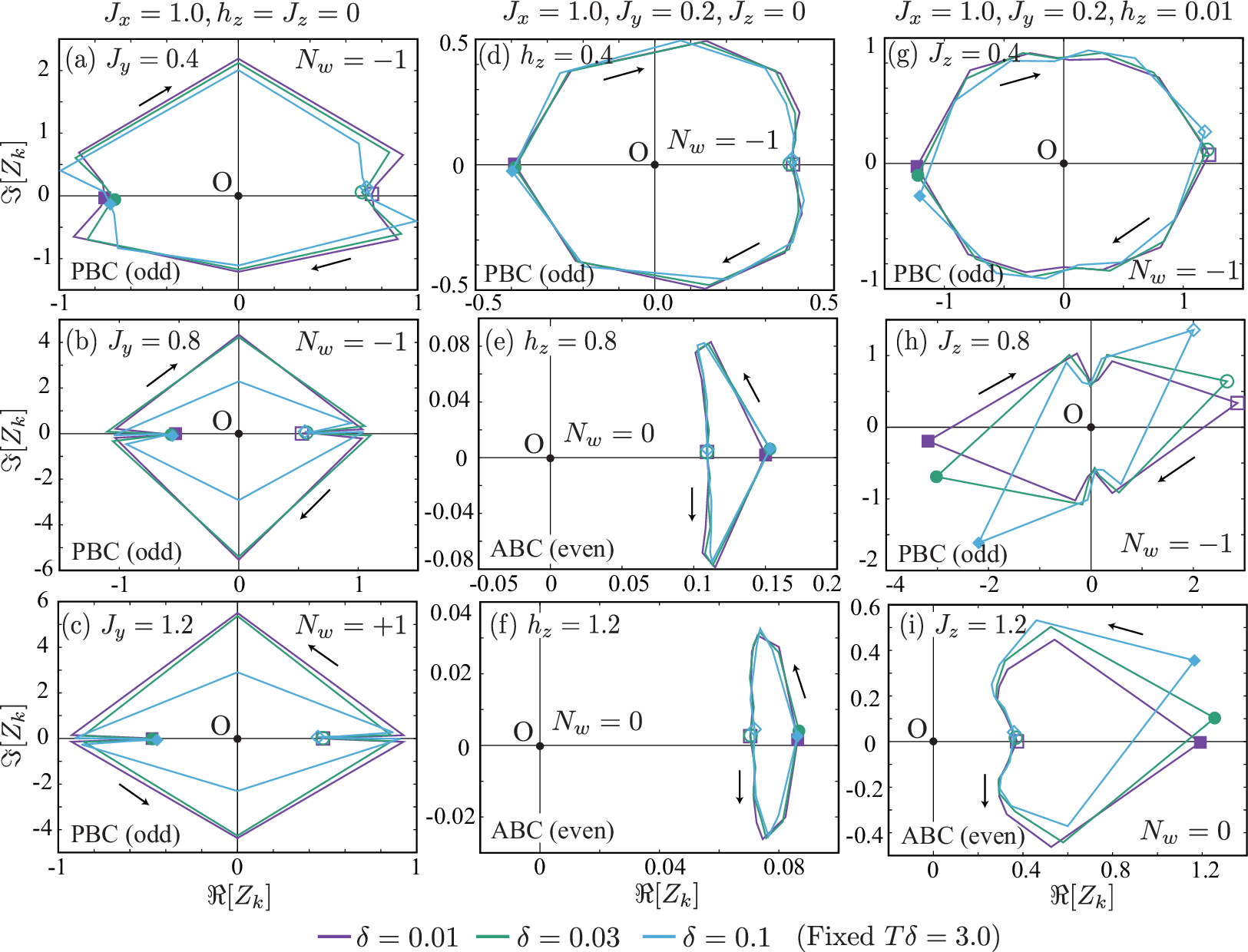}}
\caption{The MB Anderson pseudo vector $Z_k$ in the complex plane, obtained by the QC algorithm in \figref{top} for the 12-site XYZ spin chain. Each panel shows $Z_k$ for various model-parameter points, corresponding to \figref{den}. According to the ED calculations in \figref{ed} and the sufficiently-converged parity-selected VQE calculations in \figref{den}, the real GSs belong to the odd parity subspace for (a-d,g-h) and  the even parity subspace for (e-f,i). The BC for the XYZ spin chain is PBC (ABC) with the odd (even) parity, corresponding to the Kitaev chain with PBC (see \tabref{BC}).  Purple, green, and cyan solid lines represent the damping factors $\del=0.01$, $0.03$, and $0.10$, respectively, with fixed $T\del=3.0$ and $\Del t=0.01$. Open (close) symbols denote $Z_{k=0}$ ($Z_{k=\pi}$), and arrows indicate the ascending order of $k$. The MB winding number $N_w$ is defined by the number of times that the MB Anderson pseudo vector $Z_k$ circulates counterclockwise around the origin.}
\label{fig:zk}
\end{figure*}

The real part of the transition amplitude in \equref{TEGF} corresponds to the expectation value of the $x$ component of Pauli matrix of an ancila qubit $\sig_a^x$ based on the Hadamard test as follows,
\begin{equation}
\Re \braket{\gam_j^s(t)|\gam_{j^\pri}^a(t)}=\bra{\psi_{j,j^\pri}(t)}\sig_a^x\ket{\psi_{j,j^\pri}(t)}
\end{equation}
with
\begin{align}
  \label{eq:psijj}
\ket{\psi_{j,j^\pri}(t)}&=\frac{1}{\sqrt{2}}\br{\ket{1}_a\ket{\gam_j^s(t)}+\ket{0}_a\ket{\gam_{j^\pri}^a(t)}}.
\end{align}
Therefore, we can calculate the expectation value in the QC given in \figref{top}(a).
(Note that a similar technique is recently proposed, to calculate the Green functions of fermions~\cite{Endo2020}, while the parity is not considered.)
To obtain the time-evolved state, we have to adopt the real GS chosen between the GSs with different parities, $\ket{\mrm{gs}_\pm}=U_\psi^\pm\ket{\mrm{i}_\pm}$, where $U_\psi^\pm$ represents the optimized QC obtained by the parity-selected VQE method.
The MF operator $\gam_j^\tau$ ($\tau=s,a$) can be introduced by a control unitary gate [see \figref{top}(b,c)].
The time evolution of the Kitaev chain with PBC in \figref{top}(a) is given by the Trotter decomposition of the time evolution of the corresponding spin Hamiltonian $U_T^\pm(t)=\ex^{-\im\ham_\rS t}$, i.e., the product of the infinitesimal time evolution with local unitary operators \equref{u2}, 
\begin{align}
  u_{2}(\Del t) &= U_{2,a}(\tht_{a})U_{2,b}(\tht_{b})U_{2,c}(\tht_{c}),\\
  u_b^\pm(\Del t) &= U_{2,a}(\mp\tht_{a})U_{2,b}(\mp\tht_{b})U_{2,c}(\tht_{c}),\\
  u_1(\Del t) &= U_1(\vtht),
\end{align}
with
\begin{equation}
 \tht_a=\frac{J_x+J_y}{4}\Del t,\ \tht_b=\frac{J_x-J_y}{4}\Del t,\ \tht_c=\frac{J_z}{4}\Del t, \ \vtht=\frac{h_z}{2}\Del t.
\end{equation}
Among these operators, the boundary term $u_b^\pm(\Del t)$ depends on the GS parity in the XYZ spin chain (see \tabref{BC}). 
Moreover, the BC is inverted by the MF operator $\gam_{j^\pri}^a$ in the $\ket{1}_a$ subspace in \figref{top}(d) [see also \equref{TEMFES} and \equref{psijj}].

\Figref{zk} shows the MB Anderson pseudo vector $Z_k$ in the complex plane, obtained by the QC algorithm in \figref{top}, for various points in the model-parameter space of the 12-site XYZ spin chain, where the parameter points are the same as \figref{den}.
The MB winding number is defined by the number of times that the MB Anderson pseudo vector $Z_k$ circulates counterclockwise around the origin as well as the Anderson pseudo vector in the TB model.
Colors of solid lines show different damping factors, $\del=0.01$, $0.03$, and $0.10$, with fixed $T\del=3.0$.
Although the size and angle of $Z_k$ slightly changes with changing the damping factor, the shape is roughly kept, and thus the winding number is conserved even if the damping factor is not so small.
In addition, we have confirmed that the winding numbers are equal between the QC algorithm and the ED calculation.
Consequently, our QC algorithm for the topological invariant can basically be applied to the current NISQ devices with the serious limitation of coherent time, although the error mitigation techniques are necessarily required.

\section{\label{sec:MZM} Majorana zero mode}
In this section, we explain how to visualize the MZM.
The MZM is a zero-energy excitation of MFs, localized at edges of chain [see \figref{model}(b)].
We can understand the MZM if starting with the Majorana representation of the TB model with OBC, rewritten by
\begin{equation}
    \ham_\rM|_{\zeta=0,B=B^\pri=0} = -\im \bm{\gam}_s^T \mbf{H}_\rM \bm{\gam}_a
    \label{eq:TBMH}
\end{equation}
with the tridiagonal coefficient matrix
\begin{equation}
    \mbf{H}_\rM =\begin{pmatrix}
    \eta & g_- & & & \\
    g_+ & \eta & g_- & & \\
     & g_+ & \eta & \ddots & \\
      & & \ddots & \ddots & g_-\\
      & & & g_+ & \eta
    \end{pmatrix}
\end{equation}
and the vector of MFs,
\begin{equation}
 \bm{\gam}_\tau=(\gam_1^\tau, \gam_2^\tau, \cdots, \gam_L^\tau) \hspc (\tau=s,a).
\end{equation}
Diagonalization of the matrix is obtained by the singular-value decomposition, resulting in $\mbf{H}_\rM=\mbf{U}_\rM\mbf{\Lam}_\rM\mbf{V}_\rM^\dag$, with the unitary matrices $\mbf{U}_\rM$ and $\mbf{V}_\rM$, and the diagonal matrix $\mbf{\Lam}_\rM=\mrm{diag}\{\lam_1,\lam_2,\cdots,\lam_L\}$
with the ascending order singular values $\lam_l<\lam_{l+1}$.
Then, the TB Majorana Hamiltonian \equref{TBMH} reads,
\begin{equation}
    \label{eq:TBMH2}
    \ham_\rM|_{\zeta=0} = -\im \sum_l \lam_l \tilde{\gam}_l^s \tilde{\gam}_l^a
\end{equation}
with the superposition of MFs,
\begin{equation}
\tilde{\gam}_l^s = \sum_j (\mbf{U}_\rM^\dag)_{lj} \gam_j^s,\ \tilde{\gam}_l^a = \sum_j (\mbf{V}_\rM^\dag)_{lj} \gam_j^a.
\end{equation}

Since the superposed MFs also have the anti-commutation relation $\{ \tilde{\gam}_l^\tau , \tilde{\gam}_{l^\pri}^{\tau^\pri} \} = 2\del_{l,l^\pri}\del_{\tau,\tau^\pri}$ with the Hermiticity $(\gam_l^\tau)^\dag=\gam_l^\tau$ for $\tau (\tau^\pri)=s,a$, we can put one fermion on two MFs, $\tilde{\gam}_l^s = \tilde{c}_l^\dag + \tilde{c}_l$ and $\tilde{\gam}_l^a =  \im\br{\tilde{c}_l^\dag - \tilde{c}_l}$.
With these fermions, the TB Hamiltonian \equref{TBMH2} is rewritten by,
\begin{equation}
    \ham_\rM|_{\zeta=0} = -2\sum_l \lam_l \br{\tilde{c}_l^\dag \tilde{c}_l-\frac{1}{2}}.
\end{equation}
Therefore, the singular values are considered as the eigen-energies.
If there is a zero singular value $\lam_1=0$, the pair of MFs, $\tilde{\gam}_1^s$ and $\tilde{\gam}_1^a$, becomes the MZM.

\begin{figure}[tb]
\includegraphics[width=0.6\columnwidth]{\dfigs{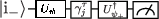}}
\caption{QC algorithm for the real-space distribution of the MZM, $|\bra{\gs_+}\gam_j^\tau\ket{\gs_-}|$ for $\tau=s,a$. The probability to observe $\ket{0}^{\otimes L}$ corresponds to the transfer amplitude $|\bra{\gs_+}\gam_j^\tau\ket{\gs_-}|$. Note that to prove it the zero-energy mode, we have to additionally check that the GS energies with different parities are degenerate utilizing the parity-selected VQE method with OBC.}
\label{fig:qc-mzm}
\end{figure}

The on-site MFs $\gam_j^\tau$ consist of creation and annihilation operators of fermion, so that single operation of the superposed MFs changes the fermion parity $\pF$.
Hence, the expectation value of the MFs for any parity-conserved eigenstates is always zero.
Instead, to visualize the MZM, we need to see the transfer amplitude without finite energy excitation between different parity subspaces, e.g., the real-space distribution for symmetric mode reads
\begin{equation}
  |\bra{\gs_+}\gam_j^s\ket{\gs_-}| = \av{\sum_l  (\mbf{U}_\rM)_{jl} \bra{\gs_+}\tilde{\gam}_l^s\ket{\gs_-}}=|(\mbf{U}_\rM)_{j1}|,
\end{equation}
because the transfer amplitude is zero except for $l=1$, namely, $|\bra{\gs_+}\tilde{\gam}_l^s\ket{\gs_-}|=\del_{l,1}$, if the first singular value is only zero, $\lam_1=0$.
Therefore, we can visualize the real-space distribution of the MZM by calculating the transfer amplitude, $|\bra{\gs_+}\gam_j^\tau\ket{\gs_-}|$ for $\tau=s,a$ in the QC.

\begin{figure*}[tb]
\includegraphics[width=1.0\textwidth]{\dfigs{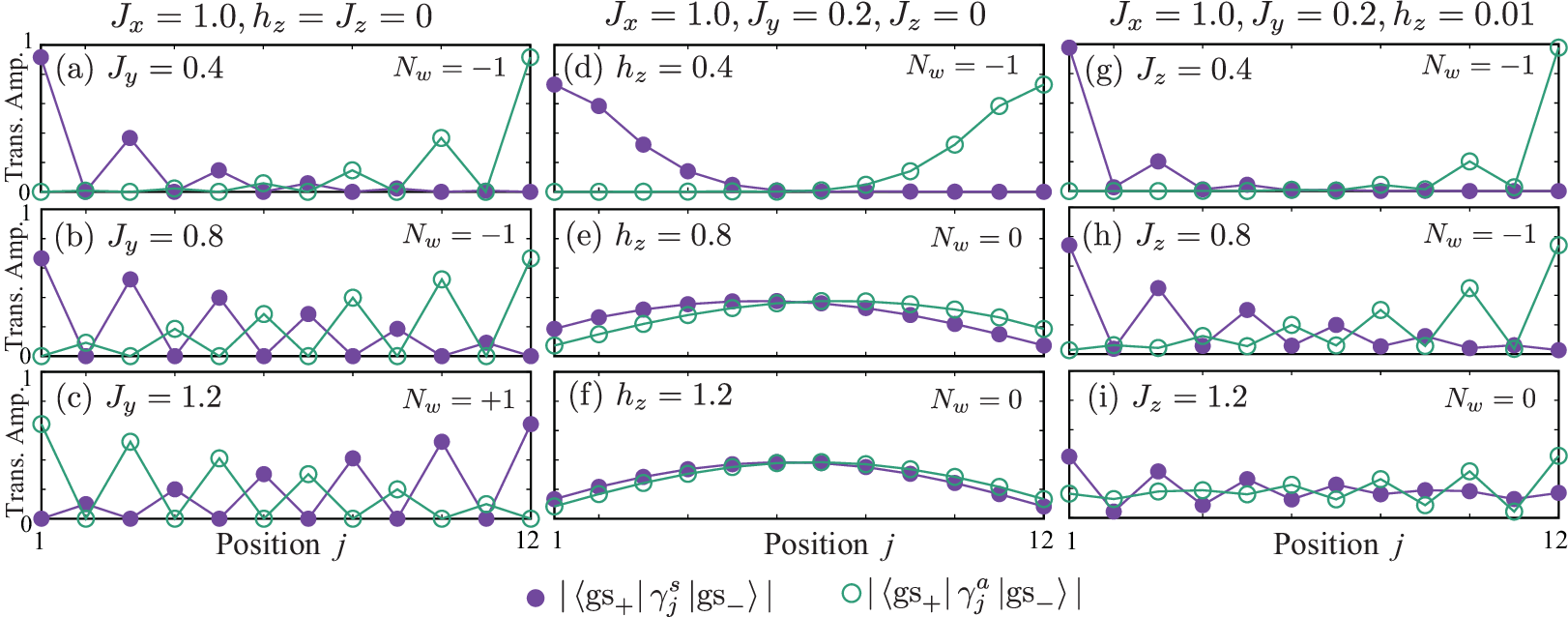}}
\caption{Transfer amplitudes $|\bra{\gs_+}\gam_j^\tau\ket{\gs_-}|$ in the 12-site XYZ spin chain utilizing the QC algorithm shown in \figref{qc-mzm} for various model-parameter points, corresponding to \figref{den} and \figref{zk}. If the GS energies of OBC with different parities are degenerate, the transfer amplitudes represent the real-space distribution of the MZMs. $N_w$ represents the MB winding number obtained in \figref{zk}}
\label{fig:mzm}
\end{figure*}

\Figref{mzm} shows the transfer amplitudes $|\bra{\gs_+}\gam_j^\tau\ket{\gs_-}|$.
The weight of MZM are localized at an edge and rapidly decreases with entering the bulk for the topological states [\figref{mzm}(a-d,g-h)].
Furthermore, we can see that the symmetric and antisymmetric modes switch the position if the winding number changes its sign [compare (b) and (c) in \figref{mzm}].
The numerical cost of this QC algorithm for the MZM is much lower than that for topological invariant, so that it is easier to confirm the topological state with the MZM visualization.
However, in this case, we should be careful with the energy difference between the GSs in even and odd parity subspaces, because there is usually an energy gap due to the finite-size effect.

\section{Summary and Discussion}
For realizing the long-term FTQC, topological states of matter are important, while the QC algorithms to determine topological invariant are still not sufficient.
In this paper, we propose the QC algorithm for topological invariant by using time evolution.
Since this algorithm requires the GS with keeping the fermion parity, we also show the parity-selected VQE method.
In addition, we propose how to visualize the MZM by the QC algorithm, and demonstrate it on the QC simulator, qulacs~\cite{Suzuki2021}, in the classical computer.

As the result of parity-selected VQE calculations, we find that the GSs with odd parity are comparably difficult to be obtained.
The non-uniform initial state before unitary operations in the QC may affect the convergence in the shallow QC.
In the QC algorithm of topological invariant, we clarify that introducing the damping factor and the cutoff time only gives a slight change in the size and angle of the MB Anderson pseudo vector, but keeps the topological invariant.
This feature guarantees the stability of our algorithms even with an inevitable noise in NISQ devices, while the shallow QC due to the short coherent time might make the topological character somewhat unstable.
Then, to execute our algorithms in NISQ devices, combining with error mitigation techniques and long time evolution algorithms will be crucial.
Alternatively, for the visualization of the MZM, our algorithm only requires the shallow QC, and thus, its demonstration is possible even in current NISQ devices.

\begin{acknowledgments}
This work was supported by MEXT Quantum Leap Flagship Program (MEXT QLEAP) Grant No.~JPMXS0118067394 and JPMXS0120319794, Grant-in-Aid for Scientific Research (B) (Grants No.~24K00586) from JSPS, Japan, JST PRESTO (Grant No. JPMJPR24F4), and the COE research grant in computational science from Hyogo Prefecture and Kobe City through Foundation for Computational Science.
Numerical computation in this work was partly carried out on the supercomputers at JAEA.
\end{acknowledgments}

\bibliography{refs-a}

\end{document}